\documentclass[preprint2]{aastex6}
\usepackage{amsmath,amstext}
\usepackage{tikz,xspace,alltt,bm}
\usetikzlibrary{shapes.geometric,arrows,positioning,calc}
\usepackage[all]{hypcap} 

\hypersetup{citecolor=blue,urlcolor=magenta,linkcolor=blue}

\newcommand\json{{\tt JSON}\xspace}

\newcommand\github{{\tt github}\xspace}
\newcommand\python{{\tt Python}\xspace}
\newcommand\scipy{{\tt SciPy}\xspace}
\newcommand\emcee{{\tt emcee}\xspace}
\newcommand\osc{{\it Open Supernova Catalog}\xspace}
\newcommand\oac{{Open Astronomy Catalog}\xspace}
\newcommand\oacs{{\oac{s}}\xspace}
\newcommand\mosfit{{\tt MOSFiT}\xspace}
\newcommand\csr{\chi_{\rm red}^2\xspace}
\newcommand\fnsep{\textsuperscript{,}}

\tikzstyle{user} = [ellipse, fill=yellow!20, text width=1.5cm, minimum width=2cm, text centered, draw=black, minimum height=1cm]
\tikzstyle{file} = [rectangle, rounded corners, text width=1.5cm, minimum width=2cm, minimum height=1cm, text centered, draw=black, fill=red!30]
\tikzstyle{web} = [trapezium, trapezium left angle=80, text width=1.5cm, trapezium stretches body, trapezium right angle=100, text centered, draw=black, fill=blue!30]
\tikzstyle{process} = [rectangle, text width=1.5cm, minimum width=2cm, minimum height=1cm, text centered, draw=black, fill=orange!30]
\tikzstyle{repo} = [diamond, text width=1cm, minimum width=1cm, minimum height=1cm, text centered, draw=black, fill=green!30]
\tikzstyle{arrow} = [thick,->,>=stealth]

\shorttitle{MOSFiT}
\shortauthors{Guillochon et al.}

\begin{document}

\title{MOSF\MakeLowercase{i}T: Modular Open-Source Fitter for Transients}
\author{James~Guillochon\altaffilmark{1}, Matt~Nicholl\altaffilmark{1}, V.~Ashley ~Villar\altaffilmark{1}, Brenna Mockler\altaffilmark{2}, Gautham~Narayan\altaffilmark{3}, Kaisey~S.~Mandel\altaffilmark{4,5}, Edo~Berger\altaffilmark{1}, Peter~K.~G.~Williams\altaffilmark{1}}
\affil{$^1$Harvard-Smithsonian Center for Astrophysics, 60 Garden St., Cambridge, MA 02138, USA\\
$^2$Department of Astronomy and Astrophysics, University of California, Santa Cruz, CA 95064, USA\\
$^3$Space Telescope Science Institute, 3700 San Martin Drive, Baltimore, MD 21218, USA\\
$^4$Institute of Astronomy and Kavli Institute for Cosmology, Madingley Road, Cambridge, CB3 0HA, UK\\
$^5$Statistical Laboratory, DPMMS, University of Cambridge, Wilberforce Road, Cambridge, CB3 0WB, UK}
\email{jguillochon@cfa.harvard.edu}

\begin{abstract}
Much of the progress made in time-domain astronomy is accomplished by relating observational multi-wavelength time series data to models derived from our understanding of physical laws. This goal is typically accomplished by dividing the task in two: collecting data (observing), and constructing models to represent that data (theorizing). Owing to the natural tendency for specialization, a disconnect can develop between the best available theories and the best available data, potentially delaying advances in our understanding new classes of transients. We introduce \mosfit: the Modular Open-Source Fitter for Transients, a \python-based package that downloads transient datasets from open online catalogs (e.g., the Open Supernova Catalog), generates Monte Carlo ensembles of semi-analytical light curve fits to those datasets and their associated Bayesian parameter posteriors, and optionally delivers the fitting results back to those same catalogs to make them available to the rest of the community. \mosfit is designed to help bridge the gap between observations and theory in time-domain astronomy; in addition to making the application of existing models and creation of new models as simple as possible, \mosfit yields statistically robust predictions for transient characteristics, with a standard output format that includes all the setup information necessary to reproduce a given result. As large-scale surveys such as LSST discover entirely new classes of transients, tools such as \mosfit will be critical for enabling rapid comparison of models against data in statistically consistent, reproducible, and scientifically beneficial ways.
\end{abstract}

\maketitle

\section{Introduction}

The study of astrophysical transients provides a unique opportunity to explore the interplay of physical laws for states of matter that are not easily reproducible in Earth-based laboratories. While the modeling of steady-state systems can yield valuable information on physics under fixed conditions, the dominance of different physical processes at different times in a given transient's evolution means that tight constraints can be placed on these processes by self-consistent modeling of their time-dependent, observable features. 

Transient characterization extends thousands of years to the first supernovae observed in antiquity, and the dataset has grown to be very rich in the past century at the same time that astronomical methods have become more rigorous. Over the past several decades, technology for collecting time-domain data has changed from predominantly photographic plates to charge-coupled devices, and the standards for characterizing the brightness and color of transients has evolved in tandem. Some of the best-characterized transients date from an era before cheap computation and storage became ubiquitous, and are often not published with enough corresponding information to enable robust reproduction of observed data by models (information such as bandset, instrument, and/or magnitude system employed for a given observation). This means that the first step to modeling a given transient may involve contacting several people involved in the original study, a process which greatly slows the rate of scientific exploration.

The complete collection of observed transient data by astronomers has grown to a level that is easily characterizable as ``big data,'' a feature that will become more pronounced in the era of large-scale all-sky surveys that will (in the case of LSST) yield $\sim 20$~TB of imaging data every day \citep{Abell:2009a}. The products that are of interest for transient modeling (primarily photometry and spectra) is however very manageable, with the total dataset presently being $\lesssim 10$~GB in size \citep{Guillochon:2017a}, small enough to fit comfortably on a modern smart phone. But while the total number of known transients is expected to grow significantly in the coming decade, the identity of many transients will likely be difficult to determine given the limited spectroscopic follow-up available to the community. This lack of identification and characterization can reduce the utility of future surveys which have the potential to increase transient populations by orders of magnitude.

Open catalogs for astronomy \citep{Rein:2012a, Guillochon:2017a, Auchettl:2017a} aim to address these issues by agglomerating and crowd-sourcing data associated with each transient from the original publications, private communications, and publicly available resources. Such catalogs enable observers to easily compare their data to previously published works, identify transients that are similar to transients in their own datasets, and combine their own data on individual events with that from other researchers.

But while the availability of time-domain data has improved significantly, publicly accessible models of transients have remained elusive \citep[we are aware of one other service that offers conditional public access to supernova models, SNAP,][]{Bayless:2017a}. Individual works have focused on small subsets of data, offering descriptions of either light curve shapes or distributions of physical parameters for a given set of transients, but the specific data products depend heavily on the scientific motivations of the study in question. At present, reproducing a given model often requires a complete rewrite of the expressions presented by the original authors who put forward the model, which means that successful models are often times those that are simplest for others to implement, as opposed to models that best reproduce the observations.

Even in the cases where data are readily available, incompleteness in how the data are presented or ingested into catalogs can lead to the propagation of errors: for example, no distinctions between upper limits and detections, or misreporting of the magnitude system used (AB or Vega). In such cases, applying a well understood model, ideally calibrated against similar transients in the literature, can help to flag up potential errors through unrealistic model parameters or unexpectedly large residuals between model and data. Therefore if models are built to interact directly with transient catalogs, they allow us to use our physical insight about the system to resolve issues of missing or incorrect metadata.

In this paper we present the Modular Open-Source Fitter for Transients (\mosfit), a \python-based package released under the permissive MIT license that yields publicly accessible and reproducible models of transients. This paper is intended to be a descriptive guide of \mosfit and its capabilities upon its initial (version 1.0) release, but for an up-to-date user guide of the code the reader should consult the online documentation\footnote{\url{http://mosfit.readthedocs.io/}}. We note that \mosfit has already been used in the astronomical literature in at least three studies \citep{Nicholl:2017a,Nicholl:2017b,Villar:2017a}.

In Section~\ref{sec:reproducibility} we describe some of the concerns about reproducibility in astronomy, and lay out the guiding principles for the \mosfit platform and how the code is designed to make time-domain science fully reproducible. Methods for inputting data into \mosfit are described in Section~\ref{sec:input}, whereas the process for defining models in the code is described in Section~\ref{sec:models}. Products of the code, and how users can share their results, are described in Section~\ref{sec:outputs}. Assessing model performance is covered in Section~\ref{sec:scoring}, concluding with a discussion of \mosfit's present-day shortcomings and future directions in Section~\ref{sec:discussion}.

\section{End-to-end Reproducibility}\label{sec:reproducibility}

It is difficult to deny the massive impact the Internet has had upon science, especially open science efforts. Not only are scientific results immediately available via a wide range of media, but the full chain of software used to produce a scientific result is becoming increasingly available, even to the point where the platforms used to run a piece of scientific software can be replicated by third parties via virtual machines \citep{Morris:2017a}. This trend solidifies scientific results by ensuring that others can reproduce them (on a wide range of platforms via continuous integration services), enables third parties to identify possible problems in the software used to produce a given result, and fosters follow-up studies that may only require minor adjustments to an existent software stack.

These trends toward open access data policies have begun to take shape in the time-domain community, although much remains to be done. For time-domain astronomy, the issue of data access takes on a critical importance as every transient is a unique event {\it whose data can only be collected once} that will, at some level of detail, differ from every other transient previously observed, a situation that is far removed from laboratory-based experiments where identical conditions can be tested repeatedly. Even if the transients themselves are almost identical, the observing conditions will almost certainly differ between transients.

For astronomy, a useful definition of a reproducible experiment is the series of steps (pipelines) required to convert raw observational inputs into scientifically-useful data products. These pipelines can in principle be re-run at a later date to ensure the data products were produced accurately, or to provide updated data products if the methods contained within the pipeline have changed and/or more input data has become available in the interim. As publications do not yet provide repository hosting, the free hosting services offered by private companies such as \github has given viable options to observers wishing to share their pipelines \citep[for a recent example see][]{Miller:2017a}. Describing a transient with a physical model can be viewed as one of the last steps in such a pipeline: once the observational data products have been produced, a piece of modeling software consumes those products and produces higher-level products of its own.

A complication is that the end-to-end pipeline, which ideally would extend from raw photon counts/images to physical parameter inferences, are distributed amongst a finite number of scientific groups that exchange the data to one another via scientific publications, private communications, or public data repositories, with no consensus on the best way to exchange such data (see Section~\ref{sec:sharing}). Often times pieces of these pipelines are simply not available to the wider community, making a result reproducible only if all pieces of the pipeline are open and/or cooperative. This issue is particular acute on the modeling end of the pipeline, with off-the-shelf modeling software only being available for the most commonly studied transients \citep[e.g., SNooPy for Ia SNe,][]{Burns:2011a}.

While making source code for a project available is one of the first steps towards enabling others to reproduce your work \citep{Baker:2016a}, true reproducibility across platforms is difficult to achieve in practice, especially for compiled code where subtle differences in compiler behavior can yield different outcomes \citep{Colonna:1996a}, particularly in chaotic systems \citep{Rein:2017a}. While some projects have undertaken heroic efforts to ensure bit-for-bit consistency across a wide range of platforms \citep[][]{Paxton:2015a}, the required labor is often infeasible for smaller projects.

For optimization and sampling where stochastic methods are employed, bit-for-bit reproducibility is less crucial, as random variations on the initial conditions should always converge to the same solution(s) anyway. Stochastic methods offer no guarantee however that they will converge in a finite time, particularly if they are prone to getting stuck in local minima, and the users of such methods should always be wary of this possibility. The determination of when an algorithm has converged to the solutions of highest likelihood can be bolstered by repeated runs of the stochastic algorithm and/or metrics for convergence that determine if the final distribution of likelihood realizations are well-mixed (see Section~\ref{sec:scoring}).

\subsection{Guiding principles and code design}\label{sec:principles}

Mindful of the issues mentioned above, the primary goal of \mosfit is to make analysis of transient data reproducible and publicly available. The \mosfit platform has been written in \python, the most flexible choice at present for open source astronomy projects given the immense amount of development on astronomy-centric packages such as {\tt astropy}, {\tt astroquery}, {\tt emcee}, and many others. Similar to the \osc, we constructed \mosfit with a set of principles to guide us when making various code design decisions. Our goals for \mosfit as a platform are:

\begin{enumerate}
\item To enable the rapid construction and modification of semi-analytical models for transients such that scientists can react swiftly to newly-discovered transients and adjust their models accordingly (or to construct entirely new models).
\item To make the ingestion of historical and contemporary observational data as painless for the user as possible, and minimizing the need for scientists to scrape, annotate, and convert data into the proper input form.
\item To provide fits of models to data that are assessed by scoring metrics that are related to the total evidence in favor of a given model (as opposed to simple goodness-of-fit tests), which have the potential to be used for model comparison.
\item To execute those models in a computationally efficient way that minimizes runtime and encourages users to optimize critical pieces of code that are likely shared by many models.
\item To provide predictions of the {\it physical} parameters responsible for an observed transient (e.g., ejecta mass or explosion energy) rather than shape parameters that are not simply relatable to physical processes.
\item To distribute the work of modeling transients amongst scientists and the public and enabling sharing of their results to the broader community.
\item Finally, to enabling sharing of user fits to transients that are publicly accessible on a rapid timescale, potentially hours after a transient's data is first made available.
\end{enumerate}

\begin{figure*}
\centering
\includegraphics[width=\linewidth]{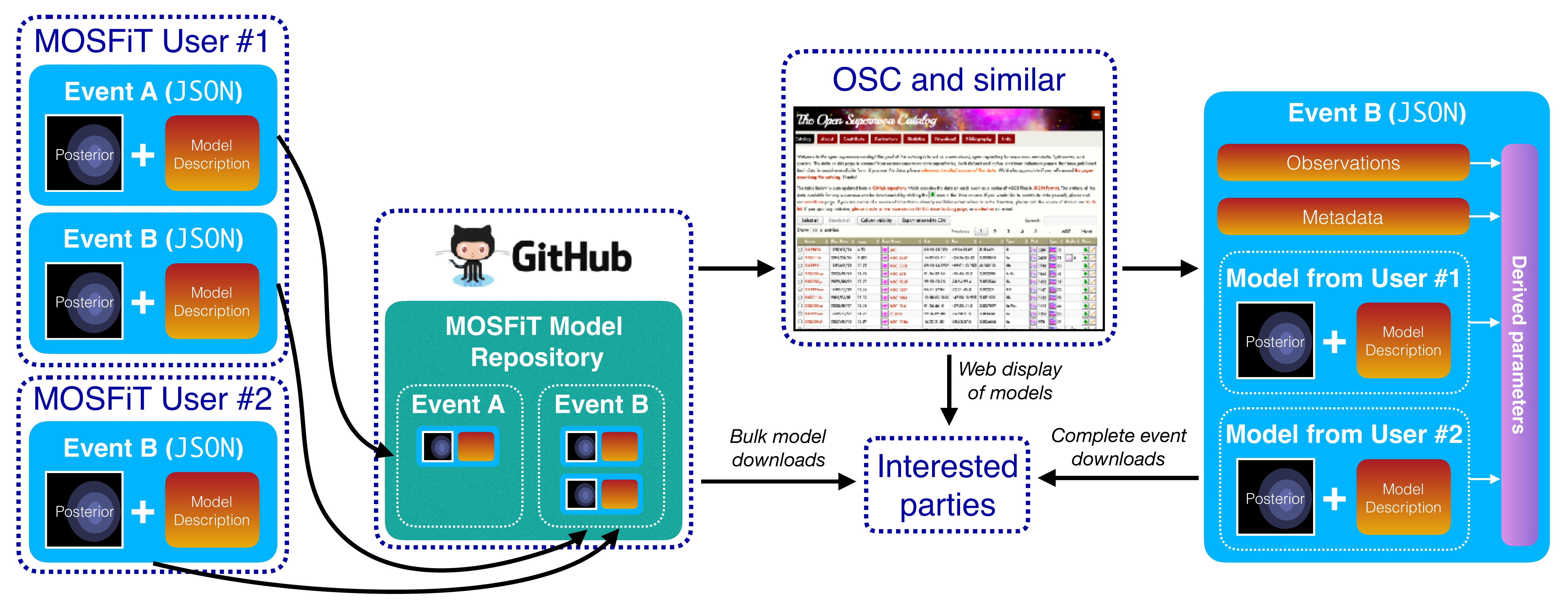}
\caption{Typical interactions of \mosfit users with the \oacs and consumers of their data products. In the above example, two \mosfit users (\#1 and \#2) submit model fits for two different events (A and B) to a model repository on \github via \mosfit's upload feature. The data in the model repository is then absorbed by the \oacs such as the OSC, which can then deliver event information to interested parties that contains observational data, model fits to that data, and any parameters derived from the model fits. The primary form of data exchange between various users and services are \json files, displayed as light blue.}
\label{fig:interactions}
\end{figure*}

More succinctly, \mosfit should be {\it easy}, {\it adaptable}, {\it fast}, {\it accurate}, {\it transparent}, and {\it community-driven}. These goals are all served by making the platform open-source, well-documented, modular, optimized, and considerate of the astronomy software ecosystem both at the present day and in the future. \mosfit is intended to be used by both observers and theorists, and so should be useful to both parties; some should be able to use \mosfit as a development platform for constructing new transient models, whereas others should simply be able to use \mosfit as a tool to match well-vetted models against new transients.

\section{Data input}\label{sec:input}

The story of how a particular photometric dataset makes its way from collection to publication differs on where the data was collected, who reduced the data, and how the data was presented. While any individual dataset is usually not too difficult to manipulate into a proper input format for a given code, the process can be extremely tedious if multiple datasets from multiple sources need to be converted. \mosfit aims to simplify this process greatly by relying upon the \oacs\footnote{\url{https://astrocats.space/}} to provide sanitized, homogeneously formatted data, which can be optionally supplemented by the user's own private data. The flow of data to and from the \oacs is shown in Figure~\ref{fig:interactions}.

Our goal is to make the default choices of algorithms employed by \mosfit robust enough such that running a model fit against new data has the best possible chance of yielding an ensemble of model parameters that best explain the data. While there are always likely to be some transients where the data is not amenable to these default choices, the platform should be expected to more often than not return a meaningful result for a wide range of possible inputs.

\vfill\null
\subsection{Using data from the \oacs}

Public data can be accessed directly by name from the command line using the data available in the \oacs. As an example, the following command will download all data for PS1-11ap and prompt the user to choose a model to fit against it:

\begin{verbatim}
mosfit -e PS1-11ap
\end{verbatim}

For the user, this eliminates a significant amount of labor that might be involved in finding all the literature on this particular transient, collecting the fittable data from those sources, and combining the data into a common format. It also means that independent users will have access to exactly the same data, which is not guaranteed in cases where the original authors need to be contacted to acquire the data: the best available data from the original authors may change after publication as reductions are refined, either by the collection of better subtraction data or improvements in the reduction pipelines, meaning that supernova data is often not entirely stable.

\subsection{Using private data and arbitrary input formats}

Ideally, part of the pipeline that processes data would yield data in a common schema that does not vary between individual observers. A file's format is one part of this schema, but a format alone is not enough, as formats do not specify key names or mandate which fields should be accompany a given type of data. The popular FITS standard\footnote{\url{https://fits.gsfc.nasa.gov/fits_documentation.html}} is an example of a format widely used for its flexibility, with users appreciating the ability to specify whatever data structure best suits their particular science need. But this flexibility comes at the cost of reproducibility, with the schema of individual FITS files often being poorly documented, making it difficult to decipher the data presented in a given FITS file.

Part of \mosfit's purpose is to assist the mission of converting the transient dataset, which is spread over tens of thousands of differently-formatted ASCII and binary files, to a single schema, which is presently defined by the \oacs\footnote{\url{https://github.com/astrocatalogs/schema}}. While many of the schema's current properties have been decided in consultation with a small group of testers, the schema is not final and is intended to be modified in response to community feedback.

As a large fraction of the available data is not in this format, a {\tt Converter} class (not itself a module as it is not required for model execution) is provided with \mosfit that will perform this conversion and feed the converted data into the {\tt Transient} module. This class has been written to read ASCII data in a large number of common formats: delimited tables, fixed-width CDS format, LaTeX tables, etc. As each table provided by a source is likely to use its own style of data presentation, the converter works through a series of logical steps to attempt to infer the table's structure, and then prompts the user with a ``choose your own adventure''-style questionnaire to determine structure details that it could not determine automatically. Once this conversion process is complete, the data is converted to Open Catalog format and fed into the {\tt Transient} module.

As an example of the conversion process, consider the following input file {\tt SN2017fake.txt} in {\tt CSV} format, which presents observations in counts rather than magnitudes:
\begin{verbatim}
time,counts,e_counts,band,telescope
54321.0,330,220,B,PS1
54322.0,1843,362,B,PS1
54323.0,2023,283,B,PS1
\end{verbatim}
The user would pass the following command to \mosfit to begin the conversion process,
\begin{verbatim}
mosfit -e SN2017fake.txt
\end{verbatim}
which would then ask the user a few additional questions about the dataset that are not discernible from the input (e.g. ``what is the source of data,'' ``what instrument was used,'' ``what is the zero point of the observations''). \mosfit would then produce a new file, {\tt SN2017fake.json}, containing the data in OAC format:
\begin{verbatim}
{
    "SN2017fake":{
        "name":"SN2017fake",
        "sources":[
            {
                "bibcode":"2017FakeJ..123..45N",
                "alias":"1"
            }
        ],
        "alias":[
            {
                "value":"SN2017fake",
                "source":"1"
            }
        ],
        "photometry":[
            {
                "time":"54321.0",
                "band":"B",
                "countrate":"330",
                "e_countrate":"220",
                "e_upper_magnitude":"0.3125",
                "magnitude":"22.95",
                "telescope":"PS1",
                "u_countrate":"s^-1",
                "u_time":"MJD",
                "upperlimit":true,
                "upperlimitsigma":"3.0",
                "zeropoint":"30.0",
                "source":"1"
            },
            {
                "time":"54322.0",
                "band":"B",
                "countrate":"1843",
                "e_countrate":"362",
                "e_lower_magnitude":"0.23725",
                "e_upper_magnitude":"0.19475",
                "magnitude":"21.836",
                "telescope":"PS1",
                "u_countrate":"s^-1",
                "u_time":"MJD",
                "zeropoint":"30.0",
                "source":"1"
            },
            {
                "time":"54323.0",
                "band":"B",
                "countrate":"2023",
                "e_countrate":"283",
                "e_lower_magnitude":"0.16375",
                "e_upper_magnitude":"0.14225",
                "magnitude":"21.735",
                "telescope":"PS1",
                "u_countrate":"s^-1",
                "u_time":"MJD",
                "zeropoint":"30.0",
                "source":"1"
            }
        ]
    }
}
\end{verbatim}
After conversion, the program will then ask the user which model they would like to fit the event with out of the list of available models. This file could now be shared with any other users of \mosfit and directly fitted by them without having to redo the conversion process, and can also optionally be uploaded to the \oacs for public use.

\subsection{Associating observations with their appropriate response functions}

An important consideration when fitting a model to data is the transformation between the photons received on the detector and the numeric quantity reported by the observer. This transformation involves convolving the spectral energy distribution incident upon the detector with a response function; for photometry this function is a photometric filter with throughput ranging from zero to one across a range of wavelengths. Ideally, the filter would be denoted by specifying its letter designation (e.g., V-band), instrument (e.g., ACS), telescope (e.g., Hubble), and photometric system (e.g., Vega), as the response even for observations using a filter with the same letter designation can differ significantly from observatory to observatory. Due to temporal variations in Earth's atmosphere, actual throughput can vary from observation to observation even with all of these pieces of information being known, but it is common practice for observations to be corrected back to ``standard'' observing conditions before being presented.

Some transients may be observed by several telescopes, each with their own unique set of filters that may or may not be readily available. The Spanish Virtual Observatory's (SVO's) filter profile service \citep{Rodrigo:2012a} goes a long way towards solving this issue by providing a database of filter response functions\footnote{\url{http://svo2.cab.inta-csic.es/theory/fps/}}. \mosfit interfaces directly with the SVO, pulling all filter response functions from the service, with associations between the functions available on the SVO and combinations of filter/instrument/telescope/system being defined in a filter rules file. In cases where a given response function is not available on the SVO, it is possible to locally define filters with throughputs as a function of wavelength provided as a separate ASCII File.

\subsection{Fitting subsets of data}

When fitting data it is often desirable to exclude certain portions of the dataset, for example to test that compatible parameters are recovered when fitting against different subsets of the data, or to exclude data that is known to not be accounted for by a given model. These exclusions can be performed in a number of simple ways by the user via command-line arguments; the user can limit the data fitted to a range of times, a select few bands/instruments/photometric systems, and/or specific sources in the literature. Alternatively, the user can exclude data by manipulating the input JSON files themselves.

As described in Section~\ref{sec:sharing}, fitting against a selected subset of the input data alters the data's hash, meaning that independent fits using the same model but different subsets of the data will be regarded as being unique upon upload. Only fits with identical model and data hashes will be directly compared by the scoring metrics described in Section~\ref{sec:scoring}.

\begin{figure*}
\centering
\includegraphics[width=\linewidth]{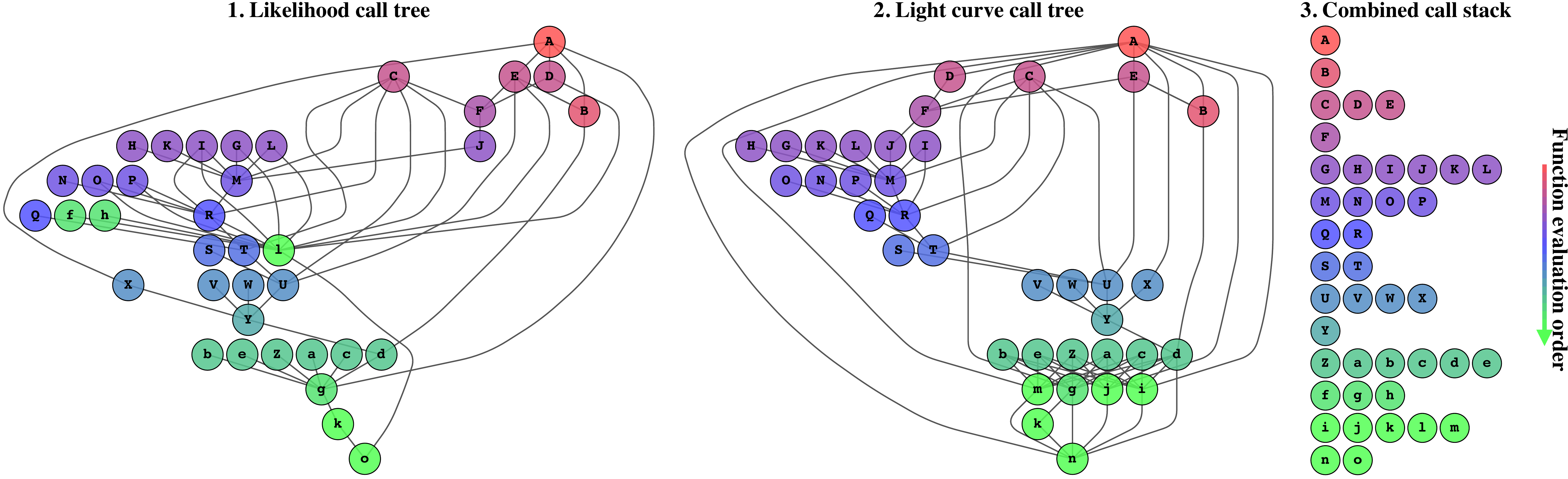}
\caption{Call trees for the (1.) likelihood and (2.) light curve functions of the superluminous supernova model as defined by its \json model files. The edges of the above graph show dependencies between the various modules (see Section~\ref{sec:modules}), which are marked here with arbitrary single letter labels. When constructing the \json files for a model, the user is responsible for specifying which modules depend on which, but is not responsible for the order in which the modules are called; \mosfit determines this order automatically, the results of which are shown in the combined call stack (3.). This ensures that even if multiple modules depend on a single module (or vice-versa) that no module is executed more than once.}\label{fig:trees}
\end{figure*}

\section{Defining models}\label{sec:models}

Each model in \mosfit is defined via two \json files, one that defines the model structure ({\tt model\_name.json}, hereafter the ``model'' file) and one that defines the parameters of the problem ({\tt parameters.json}, hereafter the ``parameter'' file). The model file defines how \python modules interact with one another to read in transient data and to produce model outputs, such as light curves and likelihood scores that are used to evaluate models. In this section, we describe generically how models are constructed out of modules, then provide a brief synopsis of the function of each of the built-in modules, and finally present the models built into \mosfit that are assembled from these modules.

\subsection{Optimal models: Constructing the call stack}

The model file defines how all of the above modules interact with one another, with each model accepting a set of inputs and producing a set of outputs that may be passed to other modules. When executing a model to produce a desired output, many of the required computations may be useful for other outputs; as an example to compute the photometry of a supernova requires one to calculate the bolometric energy inputted by its power source.

To ensure that no work is repeated, \mosfit constructs ``call trees'' (Figure~\ref{fig:trees}) that define which modules need to be chained together for a given output and generates a flat ``call stack'' that determines the order those modules should be processed, ensuring that each module is only called once. If multiple outputs are desired, multiple call stacks are constructed and combined into a single call stack, ensuring that a minimal amount of computation time is expended in producing the outputs.

\subsection{Built-in modules}\label{sec:modules}

For the optimization process above to be worthwhile, the individual modules that comprise the model must themselves be optimally written for speed and accuracy. This motivates development upon a core of built-in \mosfit modules that are general enough to be used in a variety of transient models \citep[this emulates the approach in other sub-fields of astronomy such as cosmology,][]{Zuntz:2015a}. Each module defines a single \python class (which may inherit from another class) that performs a particular function, and are grouped into subdirectories within the {\tt modules} directory depending on their purpose. These groupings are:

\begin{table*}
\centering
\caption{Table of models currently available in \mosfit.}
\label{tab:models}
\begin{tabular}{llll}
\hline\hline
Model name     & Description                           & Applicable types      & Reference(s)\\
\hline
{\tt default}  & Nickel-cobalt decay                   & Ia, Ic, PISN, Ca-rich & \citet{Nadyozhin:1994a}\\
{\tt csm}      & Interacting CSM-SNe                   & SLSN-II, IIn, ILOT    & \citet{Chatzopoulos:2013a,Villar:2017a}\\
{\tt csmni}    & CSM + NiCo decay                      & SLSN-II               & See {\tt default} \& {\tt csm}\\
{\tt exppow}   & Exponential rise, power law decay     & Any                   & \\
{\tt ia}       & NiCo decay + I-band feature           & Ia                    & See {\tt default}\\
{\tt ic}       & NiCo decay + synchrotron              & Ic                    & See {\tt default}\\
{\tt magnetar} & Magnetar engine w/ simple SED         & SLSN-I                & \citet{Nicholl:2017b}\\
{\tt magni}    & Magentar + NiCo decay                 & SLSN-I                & \citet{Nicholl:2017b}\\
{\tt rprocess} & r-process decay                       & Kilonova              & \citet{Metzger:2017b,Villar:2017a}\\
{\tt kilonova} & Multi-component r-process             & Kilonova              & \citet{Villar:2017b}\\
{\tt slsn}     & Magnetar + modified SED + constraints & SLSN-I                & \citet{Nicholl:2017b}\\
{\tt tde}      & Tidal disruption events               & TDE                   & \citet{Mockler:2018a}\\
\hline
\end{tabular}
\end{table*}

\begin{itemize}
\item {\bf Arrays}: Specialty data structures for storing vectors and matrices that are used by other modules. Examples include arrays designed to store times of observation and the kernel used for Gaussian Processes (see Section~\ref{sec:gp}).
\item {\bf Constraints}: Penalizing factors applied to models when combinations of parameters enter into disallowed portions of parameter space. An example constraint would be when the kinetic energy of a supernova exceeds the total energy input up to that time.
\item {\bf Data}: Modules that import data from external sources. At present this grouping contains a single {\tt Transient} module that is used to read in data provided in \oac format.
\item {\bf Energetics}: Transforms of the energetics into other parameters of interest, for example the velocity of the ejecta in a supernova.
\item {\bf Engines}: Energy injected by a physical process in a given transient. Examples included the decay of Nickel and Cobalt in a thermonuclear supernova, or the fallback of debris onto a black hole following the tidal disruption of a star.
\item {\bf Objectives}: Metrics used to score the performance of a given model as matched to an observed dataset. A typical choice is the ``likelihood'' of a model, the probability density of the observed data given the prediction of the model as a function of the parameters (see Section~\ref{sec:scoring}).
\item {\bf Observables}: Mock observations associated with a given transient that could be matched against collected observations. Currently only photometry is implemented, but in principle other observables (such as spectra) can be compared to observed data.
\item {\bf Outputs}: Processes model outputs for the purpose of returning results to the user, writing to disk, or uploading to the \oacs.
\item {\bf Parameters}: Defines free and fixed parameters, their ranges, and functional form of their priors.
\item {\bf Photospheres}: Description of the surface of the transient where the optical depth drops below unity and will yield photons that will propagate to the observer. These modules yield the broad properties of the photosphere(s) of the transient.
\item {\bf SEDs}: Spectral energy distribution produced by a given component. A simple blackbody is a common assumption, but modified blackbodies and sums of blackbodies, or SEDs built from template spectra, can be yielded by these routines (at present, only simple and modified blackbodies are implemented). Extinction corrections from the host galaxy and the Milky Way are also applied here.
\item {\bf Transforms}: Temporal transformations of other functions of time yielded by a given component of a transient (the central engine, an intermediate reprocessing zone, etc.), examples include reprocessing of the input luminosity through photon diffusion, or a viscous delay in the accretion of matter onto a central black hole.
\item {\bf Utilities}: Miscellaneous operators that don't fall into the above categories. Examples include arithmetic operations upon the outputs from multiple input modules, which would be used for example to sum the energetic inputs of a magnetar and the decay of radioactive isotopes in a transient where both sources of energy are important.
\end{itemize}

\subsection{Built-in models}\label{sec:builtin}

Using the modules described above, a number of transient models are constructed and included by default with \mosfit (see Table~\ref{tab:models}). While several of the models are good matches to the observed classes they represent and have been extensively tested against data, speed considerations mandate that the models not be overly complex, with simple one-zone models representing many of transients. Other models (such as the Ia model) serve as placeholders that only reproduce a given transient class' basic properties, as specialty software exists for these transients that are superior to \mosfit's model representations. For such transients it is likely that leveraging the collection of spectra on the \oacs could yield better model matches, a feature we expect to add in future versions of the code (see Section~\ref{sec:future}).

\subsection{Modifying and creating models}

In Table~\ref{tab:models}, some of the models are combinations of two models (e.g., {\tt csmni}) or simple additions to an existing model (e.g., {\tt ic}). These models share much of the code and setup of the parent models from which they inherit, and indeed creating them often only involving proper modification of the appropriate \json file.

\begin{figure}
\centering
Model A (Superluminous supernova)\\
\includegraphics[width=\linewidth]{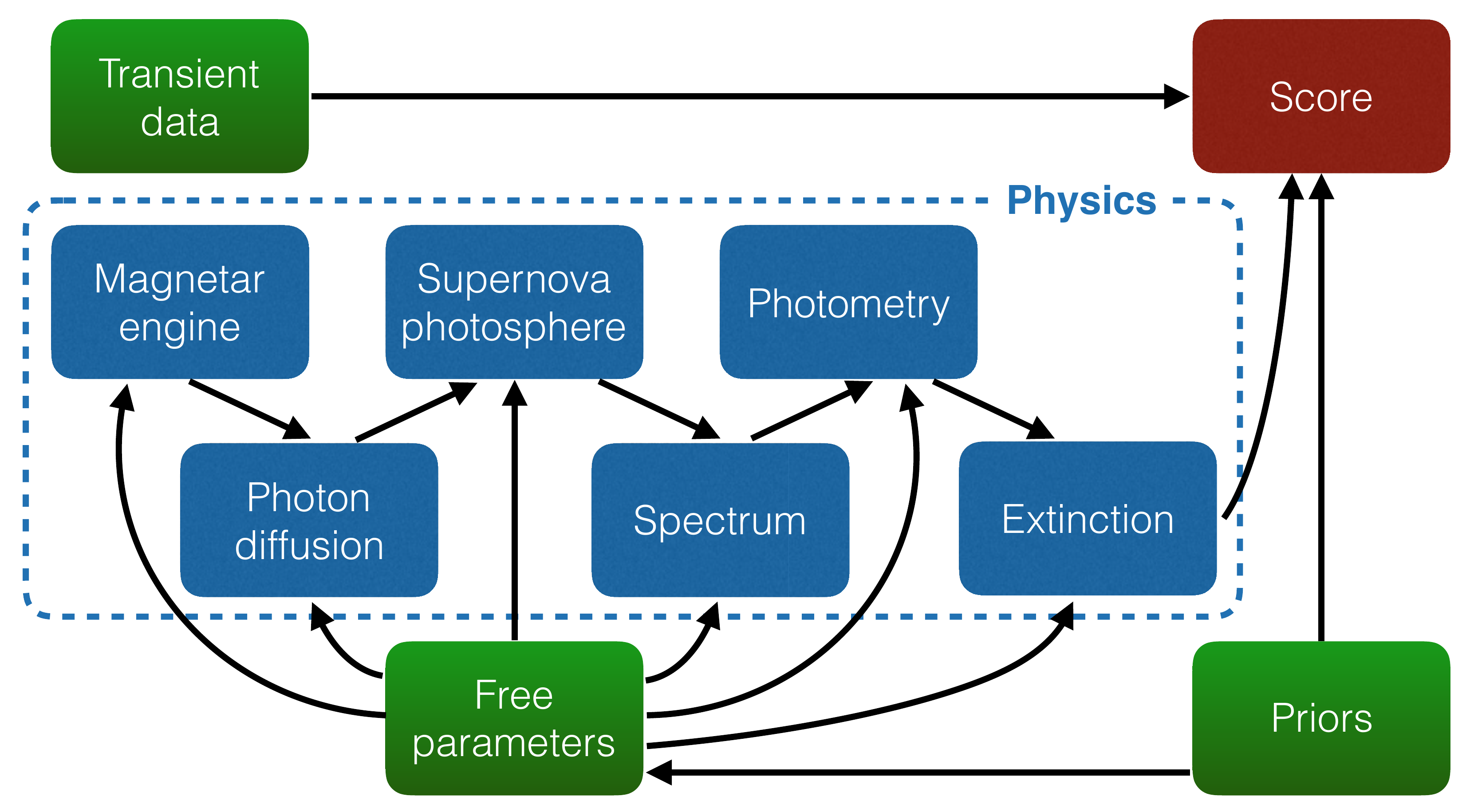}\\
\vspace{1em}
Model B (Kilonova)\\
\includegraphics[width=\linewidth]{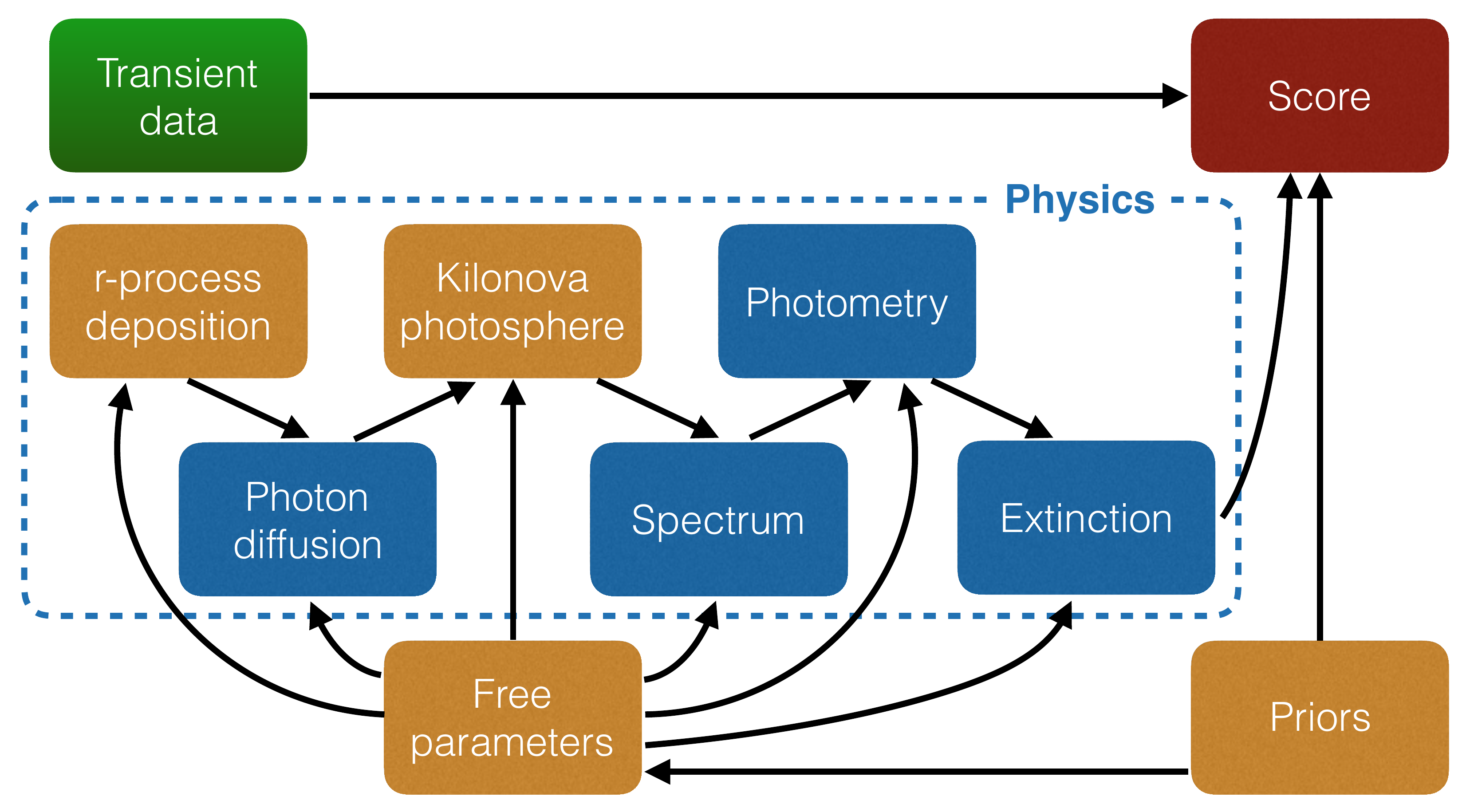}
\caption{Simplified schematic of two model trees constructed in \mosfit (not all modules shown, see Figure~\ref{fig:trees} for an example of a full tree). The top model (Model A) shows a collection of modules appropriate for describing a superluminous supernova model, whereas the bottom panel shows a model appropriate for a kilonova, constructed by replacing modules in the superluminous model (replaced modules shown in orange). In the schematic, green modules are inputs, blue modules process data from inputs and other modules, red modules are outputs, and the arrows connecting them indicate data exchange.}
\label{fig:models}
\end{figure}

The first modification a user may wish to make is altering the priors on the given free parameters of a model. By modifying the prior class in the parameters \json file, a user for example might swap a flat prior in a parameter,
\begin{verbatim}
{
    ...
    "vejecta":{
        "min_value":5.0e3,
        "max_value":2.0e4
    },
    ...
}
\end{verbatim}
for a Gaussian prior provided by a separate observation,
\begin{verbatim}
{
    ...
    "vejecta":{
        "min_value":5.0e3,
        "max_value":2.0e4,
        "class":"gaussian",
        "mu":1.0e4,
        "sigma":1.0e3
    },
    ...
}
\end{verbatim}

Next, a user might consider altering which modules are executed in a given model, for example a user might wish to switch from a simple blackbody SED to a custom SED that better describes a transient's spectral properties (as schematically shown in Figure~\ref{fig:models}). In this case, the user swaps a module (or modules) in the call stack in the model \json file, in this example from a blackbody
\begin{verbatim}
{
    ...
    "blackbody":{
        "kind":"sed",
        "inputs":[
            "texplosion",
            "redshift",
            "densecore"
        ],
        "requests":{
            "band_wave_ranges": "photometry"
        }
    },
    "losextinction":{
        "kind":"sed",
        "inputs":[
            "blackbody",
            "nhhost",
            "rvhost",
            "ebv"
        ],
        "requests":{
            "band_wave_ranges": "photometry"
        }
    },
    ...
}
\end{verbatim}
to a custom SED function with a blackbody cutoff,
\begin{verbatim}
{
    ...
    "blackbody_cutoff":{
        "kind":"sed",
        "inputs":[
            "texplosion",
            "redshift",
            "temperature_floor",
            "cutoff_wavelength"
        ],
        "requests":{
            "band_wave_ranges": "photometry"
        }
    },
    "losextinction":{
        "kind":"sed",
        "inputs":[
            "blackbody_cutoff",
            "nhhost",
            "rvhost",
            "ebv"
        ],
        "requests":{
            "band_wave_ranges": "photometry"
        }
    },
    ...
}
\end{verbatim}
Note that in the above example the name of the module and everything that calls it (in this case just the {\tt losextinction} module) was altered to accommodate the new function.

Lastly, a user may find that none of the modules available presently in \mosfit are adequate for their needs, for example if they wish to experiment with a new power source for a transient, and will create new ones to address them. In this case they can encode the required physics in a new \python class in the appropriate grouping within the modules directory (see Section~\ref{sec:builtin}), and add this module and its associated free parameters to the two \json files defining their model.

Writing new code for \mosfit requires the most care on the part of the user, as they must be mindful that the sampling and optimization routines will always be limited by the execution time of a single model realization. The models that ship with \mosfit are all computationally simple and have sub-second execution times; more complicated models that may involve integrations of systems of differential equations that may take minutes to execute per realization and thus will take that much longer when run within the \mosfit framework.

\subsection{Making models available to the community}

If a user wishes to share their model with a broader audience, the proper way to do so is to fork the \mosfit project, add their model and any supporting code, and submit that as a pull request. In general, it is the desire of the authors of this work that models contributed adhere to the following guidelines:

\begin{enumerate}
\item Parameters are preferred to be correspondent to physical properties of the transient, i.e. parameters like ejecta mass versus parameters like post-peak magnitude decline rate, although non-physical parameters are sometimes appropriate for difficult-to-describe phenomena such as spectral line features.
\item Permit broad priors on their input parameters that remain physically reasonable to support the broadest range of transients possible. Ideally, models should be capable of being applied to a broad range of transients, many of which they may fit poorly, and should extend beyond the present observed range of phenomenology to accommodate newly discovered extremal events. If a given combination of parameters is known to be unphysical, the models should penalize those combinations via constraints (see Section~\ref{sec:modules}) rather than via narrow priors that might also excluded allowed portions of parameter space.
\item Models should utilize as much of the pre-existing modules as possible as opposed to creating their own separate stack of modules that they depend on. This reduces the number of unique points of failure for individual models.
\end{enumerate}

By following these guidelines, we hope that models can be largely used ``off the shelf'' without modification by the user, which means that a larger proportion of the provided model fits will originate from the same unique models that can be more directly compared, where model uniqueness is assessed as described in Section~\ref{sec:unique}. Exact adherence to these guidelines is {\it not} mandatory, and we are open to including models that may not fit exactly into our preferred mold.

\section{Interpreting outputs}\label{sec:outputs}

The way data is ingested by a program is only half of the way we interact with software, equally important is the way that software outputs its data and the ways that output can be used. In this section we describe some of the features \mosfit provides to make its outputs as useful to the user as possible.

\subsection{Sharing fits}\label{sec:sharing}

As data reduction software evolves, and scientists move between institutions, sometimes original data can slip through the cracks. In order to preserve important research products, it is critical that data be shared in a way that it remains available indefinitely beyond its production date. The sharing of observational transient data has become increasingly common with public data repositories provided by space agencies (e.g. MAST, ESO), observing groups \citep[e.g. the CfA, SNDB, ][]{Silverman:2012a}, and third-party agglomerators \citep[e.g. WISeREP, the OSC, SNaX,][]{Yaron:2012a, Guillochon:2017a, Ross:2017a}. But for models, the means to share results is extremely haphazard, with no standard mechanism for doing so. 

\begin{figure}
\centering
\includegraphics[width=\linewidth]{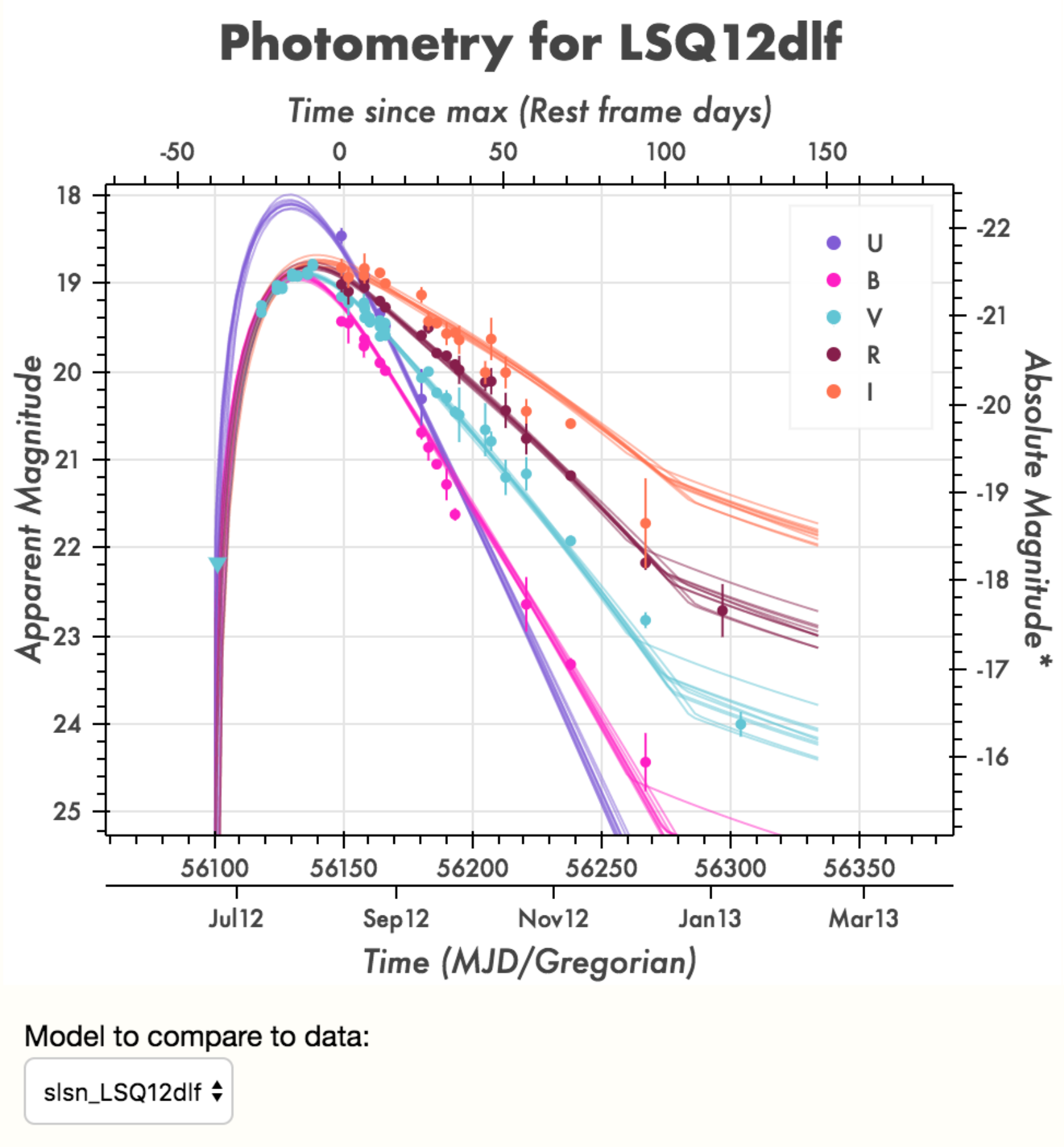}
\caption{Screenshot of figure presented on individual event page for supernova LSQ12dlf (\url{https://sne.space/sne/LSQ12dlf/}). The observed data, shown by points, is displayed alongside Monte Carlo realizations of the light curve produced by a \mosfit~{\tt slsn} run published in \citet{Nicholl:2017b}. A drop-down menu allows the user to select and display other model fits to the transient.}
\label{fig:opencatalog}
\end{figure}

In conjunction with the public release of the \mosfit software, the authors have extended the functionality of the \oacs to accommodate model fits. In addition to the observational data, the data presented on the \oacs now contain the full model descriptions in \mosfit's model format, the parameter combinations associated with each Monte Carlo realization, and light curves for all realizations, which are visually accessible on each modeled event's page as shown in Figure~\ref{fig:opencatalog}. This data can be directly loaded by the user back into \mosfit, where the user can rapidly reproduce the light curves with a different cadence, set of photometric bands, or variations on the inferred parameters (see Section~\ref{sec:synthetic}).

\subsection{Model uniqueness}\label{sec:unique}

A model can only be expected to be exactly reproducible if the code and data used to generate its output is identical. Because observing groups tend to operate independently from one another, the data fitted against from group to group is different in its time and frequency coverage, which can lead to different outcomes for parameter inferences and forward modeling. Differences in models used, even at the implementation level, such as the way an integration is performed, can also lead to real differences in outcome. Modest edits to model inputs such as changing the bounding range for a free parameter can also impact the score for a given model (or any proxy for it such as the WAIC, see Section~\ref{sec:evidence}).

To ensure that users are comparing identical models to one another, \mosfit calculates three hashes for each fit before it is uploaded to the \oacs: a hash of the input data, a hash of the model dictionary, and a hash of the \python code invoked by that model. The hashes are generated by serializing the \json/\python files associated with the input/model/code into strings, which are then passed to the {\tt sha512} algorithm which generates the hash, of which the first 16 characters are stored. This means that any change to the input/model/code will result in a different hash output, which can be used to ensure that the same data and code were used to analyze a given event. As a simple example, consider the following event with a single observation,
\begin{verbatim}
{
    "photometry":[
        {
            "time":"55123.0",
            "magnitude":"13.63",
            "band":"V"
        }
    ]
}
\end{verbatim}
which yields a hash value of {\tt 1612F22510D5A407}. Now assume that the photometry was later re-reduced and the magnitude has changed,
\begin{alltt}
\{
    "photometry":[
        \{
            "time":"55123.0",
            "magnitude":\textbf{"13.47"},
            "band":"V"
        \}
    ]
\}
\end{alltt}
this new data has a completely unique hash relative to the first, {\tt 6B59BB401C31D86D}. Together, these hashes help to reassure the user that the data and the model used to fit it are identical to what might have been produced by other users, and prevent inadvertent cross-model comparisons. One remaining reproducibility concern that these hashes do not address are changes to the external packages that \mosfit depends on, such as the outputs of various \scipy routines that may vary with \scipy version.

\subsection{Choosing a Sampler and a Minimizer}

The biggest issue in Monte Carlo approaches is convergence; as these methods are stochastic, there is absolutely no guarantee that they will ever find the best solutions, nor properly describe the distributions of the posteriors, in the time allotted to them. \mosfit embraces a ``grab-bag'' approach of techniques to maximize the chances of a converged solution.

Because the modeling in \mosfit is geared towards physical models of transients, as opposed to empirically-driven models, the evaluation of even simplistic semi-analytical models often requires the evaluation of multiple levels of non-algebraic expressions. Whereas empirically-driven models are free to choose arbitrary analytical constructions (e.g. spline fits, combinations of power laws) that are easily differentiable, purely algebraic representations of physical models are not usually possible. This means that any derivative expressions potentially required by the sampler/minimizer need to be constructed numerically. So long as the likelihood function is smooth and continuous, these derivative can be approximated via finite differencing, but this is prone to error that can make methods that assume certain constants of motion (e.g. Hamiltonian Monte Carlo, HMC) to fail to converge to the true global minimum and/or posterior \citep{Betancourt:2017a}. The rewards however are great if one is able to construct one's problem into a framework where the derivatives can be calculated in such a way that derivatives are well-behaved \citep[e.g.][]{Sanders:2015a, Sanders:2015b}, which can yield performance that scales impressively even for problems with thousands of dimensions.

\begin{figure*}
\centering
\includegraphics[width=\linewidth]{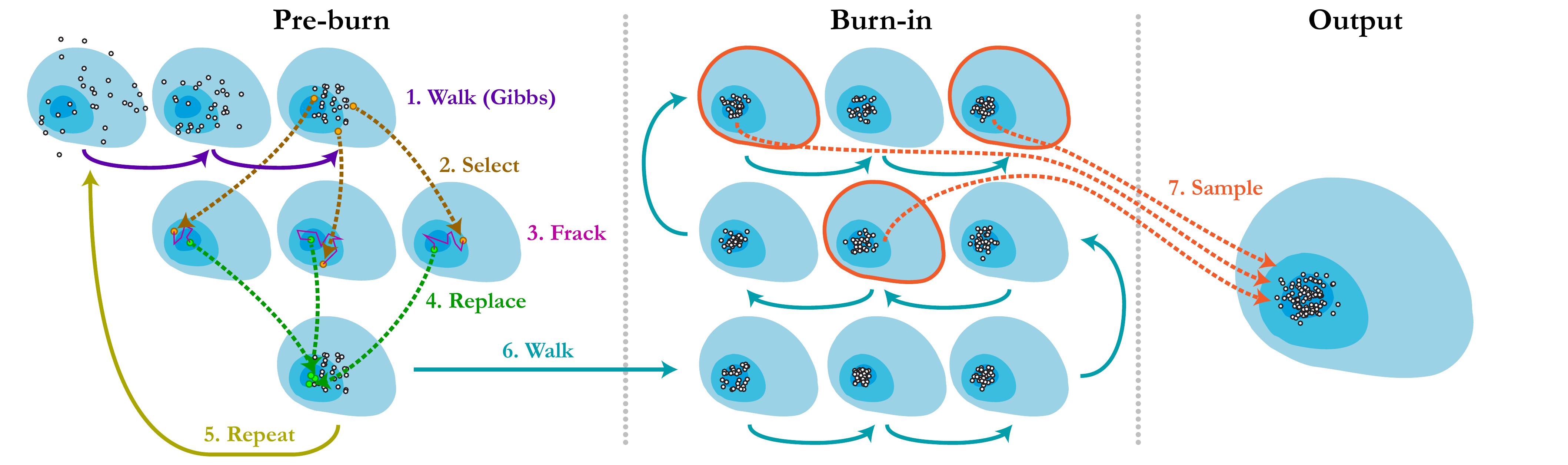}
\caption{Schematic representation of the algorithm in \mosfit for determining the parameter posterior distributions, where the cyan-colored regions indicate probability density of the likelihood function and the white circles represent walker positions. In a primary ``pre-burn'' phase, individual parameter combinations (``walkers'') are evolved using a Gibbs-like variant of the affine-invariant algorithm of \citet{Goodman:2010a} (Step~1), with walkers being selected (Step~2) periodically for optimization using \scipy's global optimizers (Step~3), the results of which are substituted back into the walker ensemble (Step~4). This process is repeated (Step~5) for a predetermined number of cycles, after which the ensemble is evolved using the vanilla affine-invariant MCMC to ensure detailed balance (Step~6). Convergence is continuously checked using the Gelman-Rubin statistic (PSRF), which once satisfied triggers the collection of uncorrelated samples over the MC chain (Step~7), with sample frequency determined by the autocorrelation time.}
\label{fig:algorithm}
\end{figure*}

Unfortunately, little quantitative information can be gained about the transients from empirically-derived modeling alone without a concrete connection to the laws of physics. Ensemble samplers, such as the \citet{Goodman:2010a} affine-invariance algorithm implemented by {\tt emcee} \citep{Foreman-Mackey:2013a}, do not require any explicit derivative definition, and thus can be used in situations where the derivatives are not easily evaluated, and even in cases where a derivative is not even definable, as is the case for discretized parameters. However, it has been shown that such methods can take an exceedingly long time to converge to simple, well-behaved posterior functions if the number of dimensions $m$ exceeds $\sim \mathcal{O}(10)$ \citep{Huijser:2015a}. This makes the vanilla {\tt emcee} algorithm completely inappropriate for problems with $m\gg\mathcal{O}(10)$, unless the function evaluations are cheap enough to run for many thousands of steps. It also suggests that caution should be exercised when interpreting posterior distributions generated by {\tt emcee} when fitting models with $m\gtrsim \mathcal{O}(10)$.

So which sampler is appropriate for modeling transients? For modeling individual transients, the choice is in favor of ensemble-based methods for their simplicity and flexibility, as physical models of transients are often able to successfully describe their bulk properties and make useful quantitative predictions even with modest $m \sim \mathcal{O}(10)$, a regime where ensemble-based methods can converge to the true posterior in a practical length of time. For hierarchical modeling of transients \citep{Mandel:2009a,Mandel:2011a,Sanders:2015b}, which can involve thousands of free parameters, ensemble-based methods are likely not appropriate unless they are used in conjunction with other methods that improve their speed of convergence.

\subsection{\mosfit's approach}

The algorithm \mosfit uses to advance walker positions is shown in Figure~\ref{fig:algorithm}. In this first release, \mosfit uses the parallel-tempered version of \emcee as its main driver. As this algorithm has been shown to preserve detailed balance, it is the only method employed to advance walker positions during the post-burn-in phase.

As the stretch-move suffers from slow convergence to the true solution in reasonably high-dimension problems, a pre-burn phase is performed that uses a variant of \emcee with a Gibbs-like stretch-move that does {\it not} preserve detailed balance. Rather than stepping in all dimensions simultaneously, the Gibbs-like sampler at each step selects a random number of dimensions $D$ to vary, where $D \in [1 - N]$, giving it more agility in the early phases where it can be easy for the walkers to become trapped in poor local minima. Once the pre-burn phase is completed, the algorithm reverts to the vanilla ensemble algorithm, the burn-in time of which has hopefully been reduced by the pre-burn procedure.

\section{Assessing model performance}\label{sec:scoring}

Goodness-of-fit can give us valuable information on a transient's properties: it can quantitatively assess which combination of physical parameters reproduce a given event, and it can suggest to us which model is best representative of a given transient. In this section, we describe three error models: $\csr$ minimization, maximum likelihood analysis, and Gaussian processes, all of which are available in \mosfit (Gaussian process being the default).

In much of the historical transient literature, goodness-of-fit has been assessed by either least squares (in cases where measurement errors are not known) or the reduced chi-square metric, $\csr \equiv \chi^2 / N_{\rm dof}$, where
\begin{equation}
\chi^2 = \sum_{i = 1}^{o} \frac{x_i^2}{\sigma_i^2},
\label{eq:csr}
\end{equation}
with $x_i \equiv O_i - M_i$ is the difference between the $i$th observation $O_i$ and model prediction $M_i(\bm{\theta})$ respectively ($\bm{\theta}$ being the free parameters), $\sigma_i$ is the normal error of the $i{\rm th}$ observation (for reference, least squares would set $\sigma_i = 1$, i.e. observational errors are ignored), $N_{\rm dof} = o - m$ is the degrees of freedom, $o$ is the number of observations, and $m$ is the number of free parameters in the model. Because of its simplicity, $\csr$ has been a favored metric when comparing models to one another. But there is danger in its simplicity: it assumes that the errors are best represented by Gaussian distributions of uncorrelated noise, an assumption that is likely untrue for observations in magnitude space and is especially inappropriate for quantifying model errors, which are dominated by systematics. If the best possible model match yields a $\csr \gg 1$, the transient is said to be {\it underfitted} by the model, suggesting that the model is incomplete, the errors in the data underestimated, or that the transient in question is better represented by different model. If the best possible match yields $\csr \ll 1$, that particular match is {\it overfitted}, suggesting the model has parameters that tune the model outputs but are not necessarily meaningful (e.g., an ad-hoc magnitude offset parameter), or that the errors in the data are overestimated, a less likely scenario than an underestimate given many difficult to quantify sources of error.

Under the assumption that the simple error model adopted is correct, $\csr$ can be directly compared between models (or different realizations of the same model), and {\it all} models with $\csr \lesssim 1$ are acceptable matches to a given transient. This means that even for different parameter combinations of the same model that there is no one ``best'' fit to a transient, and that all fits of comparable score should be presented alongside one another to gauge a model's performance, with the frequency of a given combination depending on its likelihood: a Bayesian analysis. By considering all parameter combinations that are capable of matching a sequence of observations within a prescribed tolerance, parameter degeneracies can be identified by examining the resulting posteriors. These degeneracies can be used as a tool to determine how a model could potentially be improved: as an example, a hypothetical supernova model that finds that the progenitor mass and explosion energy parameters are strongly correlated would likely benefit from an improvement to the model, such as a more-detailed calculation of ejecta velocity based upon the star's radial density profile.

So how does one select between two physically different models if both can yield model fits with $\csr \lesssim 1$? One heuristic approach that has been frequently employed is to favor the model with the lowest $\csr$, as it has the most tolerance to future changes to a model and/or data. But, under a Bayesian interpretation (with a suitably flat prior), the parameters associated with the fits of minimal $\csr$ belong to the region of parameter space for a given model with the highest posterior probability {\it density}, even if that region of parameter space is infinitesimally small. What is desired is actually the region occupied by the majority of the probabilistic {\it mass}, which may span a much wider range of parameter combinations. Given that errors in model and data are likely underestimated, even fits that yield ``poor'' $\csr$ could still correspond to reality, a feature that must be marginalized over to correctly infer a transient's parameters.

\subsection{Identifying plausible matches}

A better solution than identifying a single best fit is for the scientist to map all parameter combinations that yield plausible fits to their data. Once this map is completed, the information content of the maps of multiple models can be compared using agreed-upon metrics. In a Bayesian analysis, we identify all parameter combinations according to their posterior probability, $p(\bm{\theta} | \bm{O}) \propto p(\bm{O} |\bm{\theta}) p(\bm{\theta})$ (where $p(\bm{\theta})$ is the prior), rather than finding a single ``best-fit'' solution by minimizing the $\chi^2$ or maximizing the likelihood $p(\bm{O} | \bm{\theta})$.

One can perform a Bayesian analysis using reported measurement errors alone, in which case the likelihood is $p(O | \bm{\theta}) \propto \exp(-\chi^2 /2)$.  However, when the minimum $\chi^2_\text{red} \gg 1$, suggesting underestimated uncertainties, it is common to adopt an error model in which an additional variance $\sigma^2$ is added to all measurement errors, representing an additional source of ``white noise.''  With this error model, the log likelihood is $p(\bm{O} | \bm{\theta}) = \sum_{i=1}^n P(O_i | \bm{\theta})$, where the likelihood of a single datum is
\begin{equation}
\log p(\bm{O} | \bm{\theta}) = -\frac{1}{2} \sum_{i=1}^n \left[\frac{x_i^2}{\sigma^2 + \sigma_i^2} + \log 2 \pi \left(\sigma^2 + \sigma_i^2\right) \right]
\label{eq:mla}
\end{equation}
where $\sigma^2$ is now included within the parameter vector $\bm{\theta}$. Because the additional variance $\sigma^2$ enters into both terms for $\log p$, increasing its value both improves and penalizes the score, resulting in a balance where the variance yielded by Bayesian analysis is the additional error required to match the given model to the transient with $\csr \simeq 1$. Setting $\sigma = 0$ recovers a pure $\csr$ minimization (Equation (\ref{eq:csr})) in \mosfit, which can be accomplished via the command line ({\tt -F variance 0}).

Is the source of this additional error from the observations, or from the model, or both? Equation~(\ref{eq:mla}) assumes the additional error is normally distributed about the measurements and/or model (the additional errors could be viewed as coming from either, or both), a situation which could arise if, for example, the measurements have overestimated signal to noise ratios. For transient observations, typical errors on measurements can vary wildly, with e.g. the best photometric measurements yielding errors at the millimag level. Aside from faint detections near the detection limit of a given instrument, the signal to noise of such observations is typically well estimated, and thus a major underestimate of normally distributed errors is unlikely. As photometry is performed relative to a set of standard stars, any additional error on top of the reported stochastic error is more likely to be systematic, and is less often estimated and/or presented in the literature.

For semi-analytical models without a stochastic component, predictions can be exact to numerical precision, and thus all model errors are systematic and depend upon the level of accuracy prescribed by the computation. This means that errors in a repeated measurement (say, observing a transient with a B-band filter) are likely to be strongly correlated over some timescale, and also correlated depending on the similarity of two observations collected at the same time (e.g., the error in simultaneous B- and g-band observations are likely to be strongly correlated). This {\it serial correlation} means that the additional error introduced by a model is poorly represented by an additional error term that is normally distributed about the model mean. An error model that is more representative of serially correlated errors is thus desirable.

\subsection{Gaussian processes}\label{sec:gp}

Error models that have become recent favorites are Gaussian processes, described in depth in \citet{Rasmussen:2006a}. Gaussian processes are a non-parametric method for fitting functions, and have broad utility in their ability to provide continuous approximations to time-series data, regardless of the underlying complexity. This makes them more amendable to solutions where the model and data can be different in a wider variety of ways, permitting wider deviance at particular times and/or particular frequencies along a given transient light curve. Gaussian processes are the default error model used in all of the transient models included with \mosfit.

Gaussian processes describe the error using a covariance matrix $\bm{K}$ which contains entries $K_{ij}$ that are populated by evaluating the kernel function for every pair of observed input coordinates $i$ and $j$, from which the likelihood is computed via the expression
\begin{equation}
\log p(\bm{O}|\bm{\theta}) = -\frac{1}{2} {\bf x}^{\rm T} K_{ij}^{-1} {\bf x} - \frac{1}{2} \log \left|K_{ij}\right| - \frac{n}{2} \log 2\pi,
\label{eq:gp}
\end{equation}
where ${\bf x}$ is the vector of differences between model predictions and observation. Note that Equation~(\ref{eq:gp}) reduces to Equation~(\ref{eq:mla}) if the off-diagonal terms in $K_{ij}$ are set to zero. A shortcut exists within \mosfit to zero out the off-diagonal terms via the command line ({\tt -F covariance}).

In \mosfit the default kernel function is the squared exponential, which is defined by two lengthscales $l_{t}$ and $l_{\lambda}$, corresponding to the time between observations and the difference in average wavelength between the filters used in those observations. With the difference in time $l_{t,ij} = t_i - t_j$ and average wavelength $l_{\lambda,ij} = \bar{\lambda}_i - \bar{\lambda}_j$ between pairs of observations, the covariance matrix resulting from the application of this kernel is
\begin{align}
K_{ij} &= \sigma^2 K_{ij,t} K_{ij,\lambda} + \sigma_i^2 \delta_{ij}\\
K_{ij,t} &= \exp \left(-\frac{l_{t,ij}^2}{2 l_{t}^2}\right)\\
K_{ij,\lambda} &= \exp \left(-\frac{l_{\lambda,ij}^2}{2 l_{\lambda}^2}\right)
\end{align}
where $\sigma^2$ is the extra variance (analogous to the variance in Equation~(\ref{eq:mla})), $\sigma_i$ is the observation error of the $i{\rm th}$ observation, $t$ is the time of observation, and $\lambda$ is the mean wavelength of the observed band. The kernel is customizable by the user via the {\tt Kernel} class, but we have found that this particular functional form works well for photometric time series. In the limit of $l_t$ and $l_\lambda$ approaching zero, the GP likelihood becomes identical to the simpler error model of Equation~(\ref{eq:mla}).

If a transient is poorly constrained, e.g. the number of observations is comparable to the number of model parameters, a bimodal distribution of solutions can be returned, where some of proposed solutions are ``model-dominated'', and some solutions are ``noise-dominated,'' where the entirety of the time evolution of a given transient is purely explained by random variation \citep[see Section~5.4.1 and Figure~5.5 of][]{Rasmussen:2006a}\footnote{see also \url{http://scikit-learn.org/stable/modules/gaussian_process.html}}. In such instances, the lengthscales will typically settle on values comparable to their upper bounds, which gives the noisy solution the latitude necessary to fit all variation over a transient's full duration while still scoring relatively well. These noise-dominated solutions can often be fixated upon by the optimizer, as the average brightness of a transient can be achieved through a wide combination of physical parameters plus a long kernel length scale. In some cases the noise-dominated solution can actually perform better than a physical model; this is highly suggestive that the physical model used is not appropriate for a given transient.

\subsection{Measure of total evidence for a model}\label{sec:evidence}
In order to evaluate the total evidence of a model with $m$ free parameters, an $m$-dimensional integral must be performed over the full parameter space where the local likelihood is evaluated at every position. Unless the likelihood function is analytic and separable, such an integration is usually impossible to perform exactly, and can be prohibitive numerically without computationally economical sampling \citep[e.g., nested sampling,][]{Skilling:2004a} or approximations \citep[e.g., variation inference,][]{Roberts:2013a}.

In ensemble Monte Carlo methods like the one employed by \emcee, entire regions of the parameter space may remain completely unexplored, particularly if those regions have a low posterior density as compared to the region surrounding the global maximum. This means that the likelihood scores returned by the algorithm at individual walker locations cannot simply be added together to determine the evidence for a model.

Instead, heuristic metrics or ``information criteria'' that correlate with the actual evidence can be used to evaluate models. These criteria typically relate the {\it distribution} of likelihood scores to the overall evidence of a model, an indication of the fractional volume occupied by the ensemble of walkers. While multiple variants of the criteria exist, a simple one to implement is the ``Watanabe-Akaike information criteria'' \citep{Watanabe:2010a, Gelman:2014a} or ``widely applicable Bayesian criteria'' (WAIC), defined as
\begin{equation}
{\rm WAIC} = \log \overline{p(\bm{O} | \bm{\theta})} - \widehat{\rm var}\left[\log p(\bm{O} | \bm{\theta})\right]
\end{equation}
where $\bar{p( \bm{O} | \bm{\theta})}$ is the posterior sample mean of the likelihood, and $\widehat{\rm var}[\log p(\bm{O} | \bm{\theta})]$ is the posterior sample variance of the log likelihood, using samples from the ensemble\footnote{Note this differs by a factor -1 from \citet{Watanabe:2010a}'s original definition.}.

\subsection{Convergence}

To be confident that a Monte Carlo algorithm has converged stably to the right solution, a convergence metric should be evaluated (and satisfied) before the evolution of a chain terminates. For ensemble-based approaches, the autocorrelation time has been suggested as a way to assess whether or not convergence has been achieved \citep{Foreman-Mackey:2013a}, and running beyond this time is a way to collect additional uncorrelated samples for the purpose of better resolving parameter posteriors.

Ideally, a metric should inform the user how far away they are from convergence in addition to letting the user know when convergence has been achieved. In our testing, the autocorrelation time algorithm that ships with {\tt emcee} ({\tt acor}) is susceptible to a number of edge cases that prevent it from executing successfully, which provides the user with no information as to how close/far they are from reaching a converged state.

Instead of using this metric, we instead rely upon the ``potential scale reduction factor'' (PSRF, signified with $\hat{R}$), also known as the Gelman-Rubin statistic \citep{Gelman:1992a}, which measures how well-mixed a set of chains is over its evolution,
\begin{align}
{\rm \hat{R}} &= \frac{N + 1}{N} \frac{\hat{\sigma}_+^2}{W} - \frac{L - 1}{NL}\\
\hat{\sigma}_+^2 &= \frac{L - 1}{L} W + \frac{B}{L}\\
\frac{B}{L} &= \frac{1}{N - 1} \sum_{j=1}^{N} \left(\bar{\theta}_{j.} - \bar{\theta}_{..}\right)^2\\
W &= \frac{1}{N (L - 1)} \sum_{j=1}^{N} \sum_{t=1}^{L} \left(\theta_{jt} - \bar{\theta}_{j.}\right)^2,
\end{align}
where $N$ is the number of walker chains, $L$ is the length of the chain, $\theta_{jt}$ is the $t$th value of a parameter in the $j$th chain, $\bar{\theta}_{j.}$ is its sample mean in the $j$th chain, $\bar{\theta}_{..}$ is its global sample mean over all chains, $B/L$ is the between-chain variance, and $W$ is the within-chain variance. To calculate the PSRF for our multi-parameter models, we use the maximum of the PSRFs computed for each parameter, meaning our performance is gauged by the parameter with the slowest convergence. As described in \citet{Brooks:1998a}, a PSRF of 1.1 strongly suggests that the Monte Carlo chain has converged to the target distribution; we use this value as \mosfit's default when running until convergence using the {\tt -R} flag. Running until convergence only guarantees that a single sampling (with size equal to $N$) can be performed, users who wish to produce more samples should use more chains or run beyond the time of convergence.

The primary advantage of the PSRF over the autocorrelation time is it is always computable, which gives the user some sense on how close a given run is to being converged. In some instances, in particular if two separate groups of walker chains are widely separated in parameter space, the PSRF may never reach the target value, this usually suggests the solution is multimodal and larger numbers of walkers should be used.

\section{Discussion}\label{sec:discussion}

\subsection{Stress Testing}

With a standardized data format for transient data, \mosfit should be capable of yielding fits to any event provided in the Open Catalog format. To test this, we ran \mosfit against the full list of SNe available on the \osc with 5+ photometric measurements ($\sim$ 16,000 SNe), and found that the code produced fits for all events without error. This demonstrated that \mosfit is robust despite the broad heterogeneity of the dataset, with the full list of SNe being composed of data constructed from observations collected from hundreds of different instruments.

\subsection{Performance}

Parameter inference where the number of parameters exceeds a few can be an expensive task, particularly when the objective function itself is expensive. In our testing, a few ten thousand iterations of \emcee are typically required to produce posterior distributions in a converged state about the global maximum in likelihood space, with roughly ten times as many walkers as free parameters being recommended. As the models currently shipping with \mosfit are mostly single-zone models with relatively cheap array operations, such runs can take anywhere from a few hours (for events with dozens of detections) to a few days (thousands of detections), with the wall time being reducible by running \mosfit in parallel. This performance is reasonable and comparable to similar Monte Carlo codes, but improvements to the core modules such as rewriting them in a compiled language could bring further performance improvements.

\subsection{Synthetic Photometry}\label{sec:synthetic}

\begin{figure*}
\centering
\includegraphics[width=0.45\linewidth]{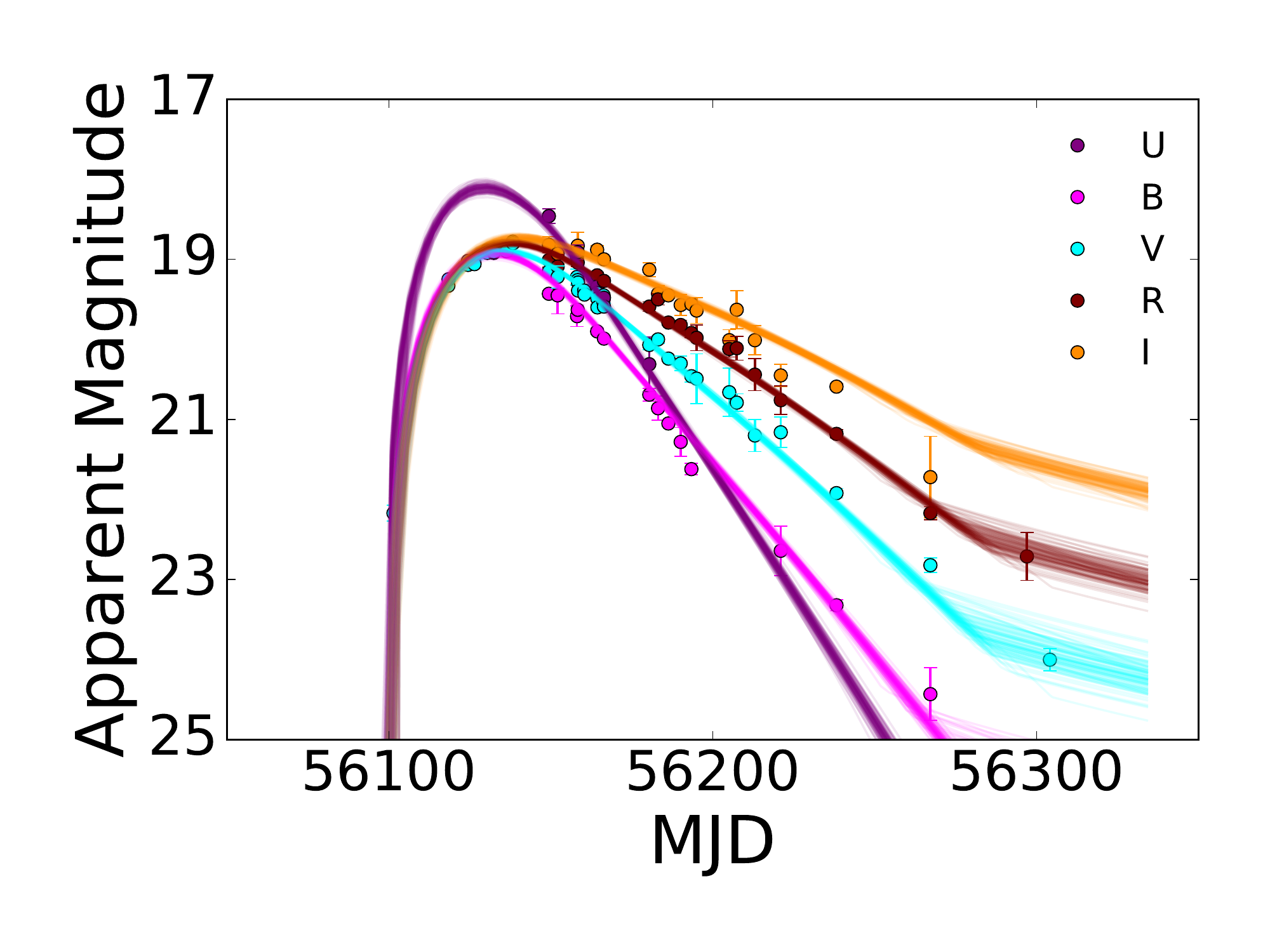}\includegraphics[width=0.45\linewidth]{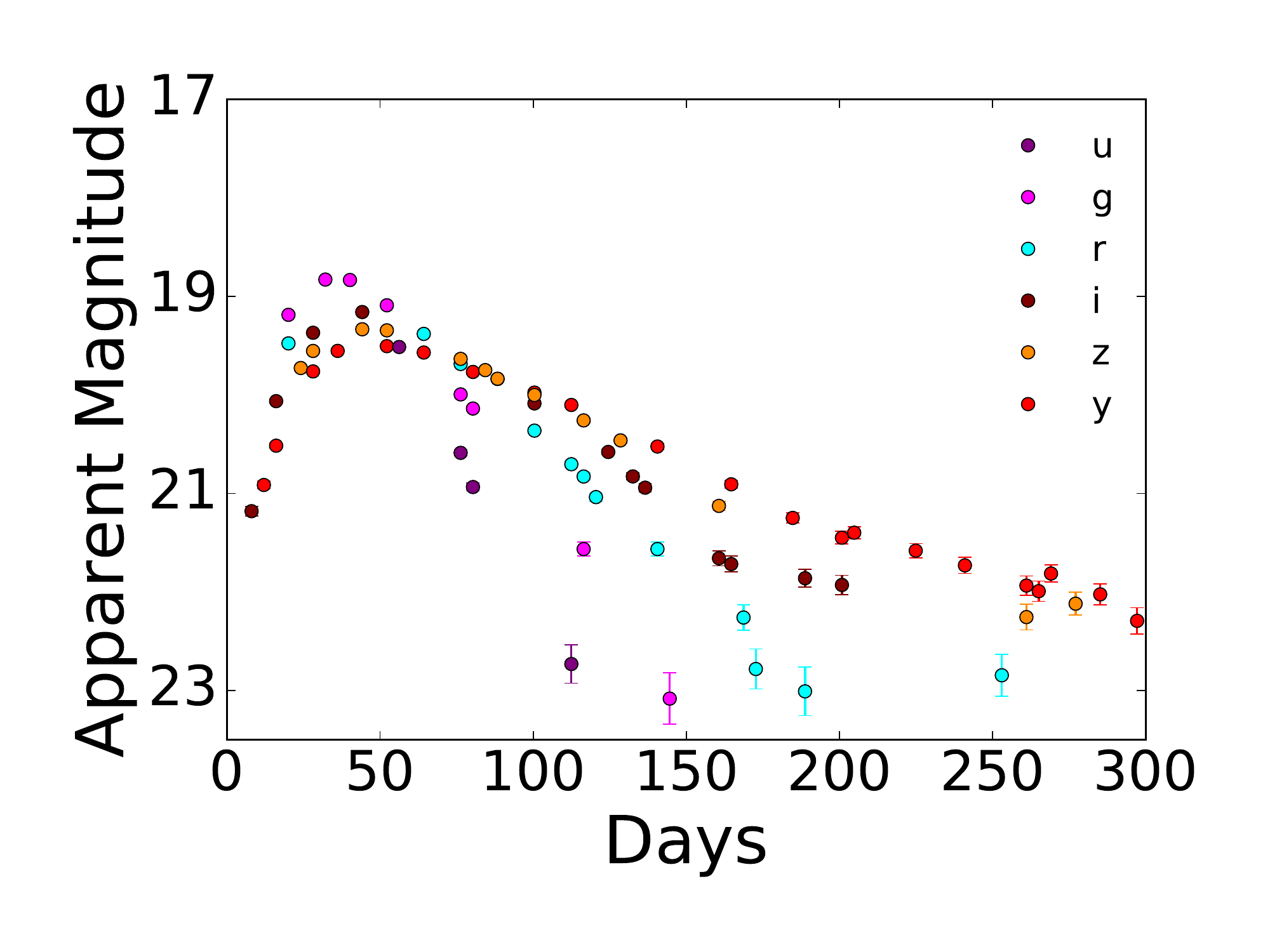}
\caption{Model matching and synthetic observations in \mosfit. In the left panel, we show the observed data for LSQ12dlf superimposed with an ensemble of models corresponding to the model posterior \citep[modeled by the {\tt slsn} model of][]{Nicholl:2017b}. In the right panel, synthetic photometry of an event similar to LSQ12dlf observed by LSST is generated by \mosfit in generative mode, where the photometry is generated by selecting a random model realization from the fits to the observed data and presuming a limiting magnitude of 22.5 in all bands. We have additionally pruned the data to roughly match the filter cadence expected from LSST.}
\label{fig:modelmock}
\end{figure*}

Once the ensembles of possible parameters have been determined for a given model, \mosfit enables the user to generate synthetic photometric observations for any instrument/band combination (Figure~\ref{fig:modelmock}). In cases where a particular transient is not being modeled, a user may wish to generate these synthetic observations from a reasonable set of priors on the physical parameters, which could then be used to produce mock observations of populations of various transients, as is done in \citet{Villar:2017a}.

Alternatively, a user may wish to instead generate a light curve based upon the results of a \mosfit run. This might be done if a user wants to supplement the model outputs with additional data perhaps not provided by the group that ran the original fit, such as the brightness in a particular band at a different set of epochs. If the model parameters and code are available, either via the \oacs or private exchange, the dataset can be loaded directly into \mosfit via the {\tt -w} flag, which loads data from a previous \mosfit run,
\begin{verbatim}
mosfit -e LSQ12dlf -w previous_run.json -i 0
\end{verbatim}
where the {\tt -i 0} flag tells \mosfit to regenerate its outputs without evolving the walker positions. Adding to this command a few additional flags (such as the {\tt -S} flag, which adds more epochs between the first/last observation, or the {\tt -E} flag, which extrapolates a number of days before/after the transient) permits the user to customize its output to their needs.

\subsection{Future versions}\label{sec:future}

While \mosfit already implements many features required of a code that ingests, processes, and produces transient data, there are several potential areas of improvement to the code that would further enhance its utility. Below, we present our feature wish-list for future releases.

\subsubsection{Flux and magnitude model matching}

A majority of the transient literature, based on historical precedent, presents observations in the form of magnitudes as opposed to fluxes. While magnitudes have the advantage of being a logarithmic scale which can better display multiple orders of magnitude of evolution in brightness, the errors in magnitude space are asymmetric and non-Gaussian, especially as the observations approach the low signal-to-noise flux limit. For simplicity, \mosfit currently yields magnitudes for the included models and compares those magnitudes to the observed values/errors, but eventually switching the model outputs to flux space would confer several advantages, including more accurate upper limits and a less approximate Gaussian process error model that could utilize symmetrical errors.

\subsubsection{Improved spectral modeling}

For transients with hotter photospheres (e.g. tidal disruption events, superluminous supernovae), the approximation of the SED as a blackbody or a sum of blackbodies still yields fairly accurate magnitude estimates for broadband filters. This assumption quickly breaks down for transients with cooler photosphere, a prime example being type Ia supernovae which have deep absorption lines even near maximum light \citep[see e.g. Figure~1 of][]{Sasdelli:2016a}, which leads to large systematic color errors.

While detailed radiative transfer models can relate parameters to output spectra \citep[e.g.][]{Botyanszki:2017a}, they are expensive and are not able to span a wide parameter space if the number of parameters exceeds a few. An alternative approach is to instead use physical parameters to predict the continuum flux (as \mosfit currently does) and to then superimpose a spectral sequence, either observed or synthetic, upon the continuum \citep[e.g.][]{Hsiao:2007a}.

\subsubsection{Survey simulations}
In the generative mode, \mosfit is able to draw light curve samples from model priors or posteriors of previous runs. These light curves can be resampled according to survey cadence and limiting magnitude to simulate observations of transient populations in a given survey (e.g. LSST). Such survey simulations could help determine the efficacy of the model comparisons we describe in Section~\ref{sec:evidence} for transients of unknown type, and could also evaluate \mosfit's utility as a real-time transient classifier. Currently, this is partially implemented into \mosfit using the {\tt --limiting-magnitude} flag, which will truncate light curves at a specified limiting magnitude. This command can be supplemented with a specification for a single cadence for all bands. More sophisticated survey simulations take into account unique filter cadences and limiting magnitudes, sky brightness and airmass as a function of location and time, and injected observational efficiencies.

\subsubsection{Flexible error models}
At the present, \mosfit includes two error models: a white noise model that adds constant variance(s) to all observations, and a Gaussian process model where the kernel is defined by two distances based on observation time and filter. The user is of course free to create new modules to implement the error model they would like to apply, but increased flexibility in the error model such as adding the option to use different families of kernels (e.g. Ornstein–Uhlenbeck, Mat\'{e}rn, etc.) would be desirable for future versions of the code. For large datasets with $O(10^4)$ points, kernel choices that lend themselves to faster inversion, as recently implemented by the code {\tt celerite} \citep{Foreman-Mackey:2017a}, would also be desirable to include as an error model option.

\subsubsection{Better estimates of marginal likelihood}
While the WAIC (Section~\ref{sec:evidence}) provides a useful heuristic for the information content of a given model fit that can be compared to other models, it is only approximate, with no measure of the error in the approximation being provided by the code. Because of this, the score can be used as a rationale to disfavor models that obviously underperform relative to others, but any rank-ordering suggested by the scores of two similarly scoring models should be done with caution. Better estimates of the information can be obtained with more walkers, but the heuristic nature of the WAIC means that its utility as a ranking mechanism is not total.

A better mechanism for computing the evidence involves substituting a different algorithm for {\tt emcee} that evaluates the evidence directly such as nested sampling, particularly dynamic sampling which typically requires far fewer function evaluations \citep{Higson:2017a}. Tests with this approach using the code {\tt dynesty}\footnote{\url{https://github.com/guillochon/MOSFiT/pull/142}}\fnsep\footnote{\url{https://github.com/joshspeagle/dynesty}} \citep{Speagle:2017a} suggests that it can yield direct evidence estimates accompanied by the measured errors of those estimates in a fraction of the function evaluations, with the side-effect of also providing far more samples of the posterior.

\subsection{Concluding remarks}

In this paper we describe the motivations behind the design of the \mosfit code and highlight its innovations in regards to data input, processing, and output. With \mosfit, we hope to make light curve analysis in time-domain astronomy more easily reproducible and accessible, but as highlighted in Section~\ref{sec:future}, many aspects of its functionality can be further improved. We aim for \mosfit to be as widely useful as possible within the time-domain community, and welcome future external contributions to the code that act to improve its performance, breadth, accuracy, and accessibility.

\acknowledgments

We thank Katie~Auchettl, Kornpob~Bhirombhakdi, Philip~Cowperthwaite, Daniel~Foreman-Mackey, Nathan~Goldbaum, Paul~``$\pi$''~Ivanov, Luke~Kelley, Ragnhild~Lunnan, Hanno~Rein, Josh~Speagle, and Lin~Yan for helpful input. The authors would also like to thank the anonymous referee for their detailed comments. This research made use of {\tt Astropy} \citep{Astropy-Collaboration:2013a}, {\tt SciPy} \citep{Jones:2001a}, and {\tt emcee} \citep{Foreman-Mackey:2013a}. K.~M. was supported in part by NSF grant AST-1516854.

\bibliographystyle{yahapj}
\bibliography{library}

\end{document}